\documentclass[12pt,a4paper]{conference}

\usepackage{fancyhdr}
\usepackage{graphicx,amsmath,amssymb,cite}
\usepackage{multind}

\newcommand{\beq}{\begin{equation}}
\newcommand{\eeq}[1]{\label{#1}\end{equation}}
\newcommand{\eeqn}{\end{equation}}

\newcommand{\beqa}{\begin{eqnarray}}
\newcommand{\eeqa}[1]{\label{#1}\end{eqnarray}}
\newcommand{\eeqan}{\end{eqnarray}}

\let\bar=\overbar

\newcommand{\Dslash}{\not{\hbox{\kern-4pt $D$}}}
\newcommand{\dslash}{\not{\hbox{\kern-2pt $\del$}}}

\newcommand{\msb}{{\bar{\ssstyle M \kern -1pt S}}}

\begin{document}

\Chapter{The hexaquark-flavoured antiK-N-N state computed microscopically with a clusterized octoquark}
           {antiK-N-N clusterized octoquark}{Bicudo, P.}
\vspace{-6 cm}\includegraphics[width=6 cm]{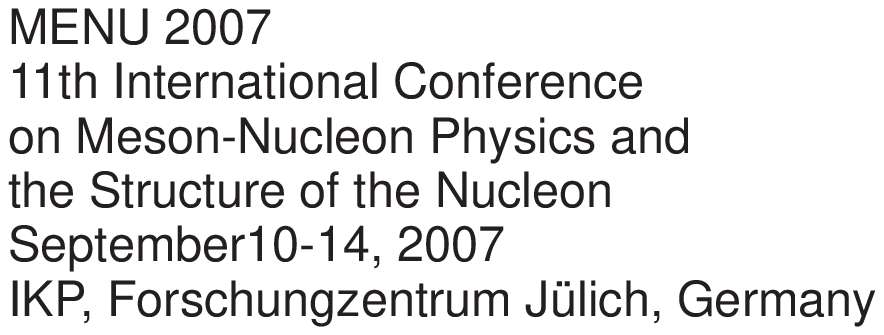}
\vspace{4 cm}

\addcontentsline{toc}{chapter}{{\it Bicudo, P.}} \label{authorStart}

\begin{raggedright}

{\it Bicudo, P. \footnote{bicudo@ist.utl.pt}}\index{author}{Bicudo, P.}\\
CFTP and Departamento de F\'{\i}sica\\
Instituto Superior T\'ecnico \\
1049-001 Lisboa\\
Portugal
\bigskip\bigskip

\end{raggedright}

\begin{center}
\textbf{Abstract}
\end{center}
The possible production processes of the antiK-NN are explored.
We derive microscopically, with the RGM, the microscopic derivation 
of the K-N and antiK-N  interactions. We discuss the binding or not
binding of the different antiK-N and antiK-NN systems. When binding
occurs, the respective decay widths are also discussed.

\section{production processes}

A possible molecular antikaon, nucleon and
nucleon three-body system e would be very interesting, both from the
exotic hadronic physics perspective 
and from the few-body nuclear physics perspective.

\begin{figure}
\begin{picture}(230,170)(0,0)
\put(40,-5){\includegraphics[width=0.75\columnwidth]{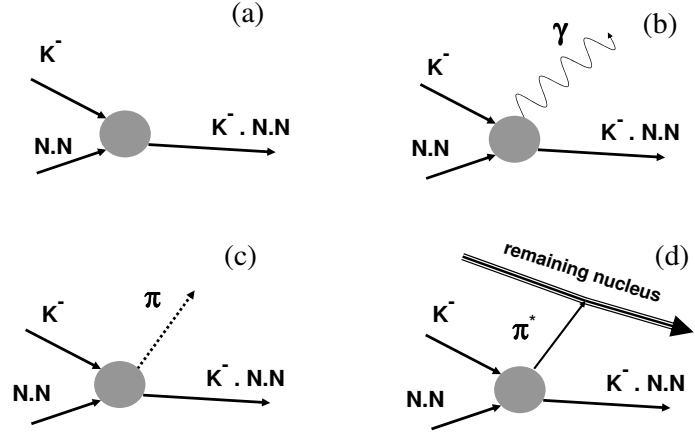}}
\end{picture}
\caption{\label{production} Different $K^- \bullet N \bullet N$ production mechanisms .}
\end{figure}

Notice that the $K^- \bullet N \bullet N$ can be produced with antikaon $(K^-)$ 
deuteron $(p \bullet n)$ scattering. Other exotic tetraquarks, pentaquarks or
hexaquarks are also 
very plausible, but they are all harder to produce experimentally because they
would need at least strangeness and charm. The several experiments dedicated to 
pentaquark searches (where not only the Kaon, but also the antikaon may interact
with nuclei), or to antikaon-nuclear binding at RCNP, JLab, KEK, DAFNE
and at many other laboratories, are already able to search for 
the proposed $K^- \bullet N \bullet N$. In particular evidence for 
$K^- \bullet N \bullet N$ has already been found by the FINUDA collaboration
at DAFNE
\cite{Agnello:2005qj}. 
In Fig. \ref{production} different possible production mechanisms are depicted. They
are similar to the $\Lambda(1405)$ production mechanisms, except that the 
$K^-$ scatters on a deuterium nucleus, not on a hydrogen nucleus. The process
in Fig. \ref{production} (a) is only possible if the width of the 
$K^- \bullet N \bullet N$ is of the order of its binding energy. The process
in Fig. \ref{production} (b) is always possible, but is suppressed by the
electromagnetic coupling. Processes in Fig. \ref{production} (c) , (d) are
dominant. 

\section{From quarks to the $K \bullet N$ }

For the $N \bullet N$ interaction we use the precise Nijmegen potentials
\cite{Stoks:1994wp}.

We compute the $K^- \bullet N$ interaction
microscopically at the quark level. 
Here we assume a standard Quark Model (QM) Hamiltonian, 
\begin{equation}
H= \sum_i T_i + \sum_{i<j\, , \, \bar i< \bar j} V_{ij} +\sum_{i \, , \, \bar j} A_{i \bar j} \ ,
\label{Hamiltonian}
\end{equation}
where each quark or antiquark has a kinetic energy $T_i$ with a
constituent quark mass, and the colour dependent two-body
interaction $V_{ij}$ includes the standard QM confining term and a
hyperfine term,
\begin{equation}
V_{ij}= \frac{-3}{16} \vec \lambda_i  \cdot  \vec \lambda_j
\left[V_{conf}(r) + V_{hyp} (r) { \vec S_i } \cdot { \vec S_j }
\right] \ . 
\label{potential}
\end{equation}
For the purpose of this paper the details of potential
(\ref{potential}) are unimportant, we only need to estimate its
matrix elements. The hadron spectrum is compatible with,
\begin{equation}
\langle V_{hyp} \rangle \simeq \frac{4}{3} \left( M_\Delta-M_N
\right) \label{hyperfine}
\end{equation}
Moreover we include in the Hamiltonian (\ref{Hamiltonian}) a
quark-antiquark annihilation potential $A_{i \bar j}$. 
Notice that the quark-antiquark annihilation is constrained 
when the quark model produces spontaneous chiral symmetry breaking.
In the $\pi$ Salpeter equation, the annihilation potential $A$ 
cancels most of the kinetic energy and
confining potential $2T+V$, 
\begin{equation}
\langle A \rangle_{S=0} \simeq 
\langle 2T+V \rangle_{S=0}
\simeq \langle V_{hyp} \rangle \ ,
\label{sum rules}
\end{equation}
leading to a massless pion in the chiral limit.
We stress that the QM of eq. (\ref{Hamiltonian}) not only
reproduces the meson and baryon spectra as quark and antiquark
bound-states,
but it also complies with the PCAC theorems.

We summarize
\cite{Bicudo:1987tz,Bicudo:1995kq,Bicudo:2004dx,Bicudo_last,Bicudo:2004cm}
the effective potentials computed for the relevant channels,
\begin{eqnarray}
V_{K \bullet N}&=& {c_K}^2
\langle V_{hyp} \rangle  
{\textstyle { 23 \over 32}\left(  1+{ 20 \over 23}  \vec 
\tau_K \cdot \vec
\tau_N \right)\ }
| \phi_{000}^\alpha \rangle \langle \phi_{000}^\alpha | \ ,
\nonumber \\
V_{\bar K \bullet N}({\mathbf r})&=& - {c_K}^2
\langle V_{hyp} \rangle  
{ \textstyle  2 \sqrt 2  
\left( 1-{ 4 \over 3}  \vec 
\tau_K \cdot \vec
\tau_N \right) }
\ e^{-{r^2 / \alpha^2}} \ ,
\nonumber \\
V_{\bar K \bullet N \leftrightarrow \pi \bullet \Lambda}&=& c_\pi c_K
\langle V_{hyp} \rangle  
{ \textstyle { 9 \over 32}\left(  1 +{ 4 \over 3}  \vec 
\tau_K \cdot \vec
\tau_N \right) }
\ | \phi_{000}^\alpha \rangle \langle \phi_{000}^\alpha | \ ,
\nonumber \\
V_{\bar K \bullet N \leftrightarrow \pi \bullet \Sigma}&=& c_\pi c_K
\langle V_{hyp} \rangle  
{ \textstyle {-5 \over 8} \left(  { 1+ \sqrt 6 \over 4} + { -3+ \sqrt 6 \over 3}  \vec 
\tau_K \cdot \vec
\tau_N \right) }\ 
\nonumber \\ && | \phi_{000}^\alpha \rangle \langle \phi_{000}^\alpha | \ ,
\label{K-N}
\end{eqnarray}
where $\vec \tau$ are $1/2$ of the Pauli isospin matrices for the 
$I=0$ and $I=1$ cases, and $c_\pi= \sqrt{E_\pi} f_\pi (\sqrt{2\pi} \alpha)^{3/2}/ \sqrt{3}$ 
is a PCAC factor, and $\mathbf r$ is the relative coordinate.
We calibrate our parameters in the two-body $K \bullet N$ channels,
where the diagonalization of the finite difference hamiltonian is straightforward.

\section{Results and conclusion}

It turns out that in the $K^- \bullet N \bullet N$,  
$I=1/2, \ S=1$ channel there is no binding
\cite{Bicudo:2007fu}. 
The groundstate has binding in the $r_{12}$ coordinate, but no binding in the
$r_{123}$ coordinate. In particular, the $r_{12}$ part of 
the wavefunction is localized and reproduces the deuteron wavefunction, while the $r_{123}$ 
part is extended over the whole size of the large box where we quantize the wave-function. 
In the limit where the size of the box is infinite, we get a bound deuteron $p \bullet n$ 
and a free $K^-$.

In the $S=0$, $K^- \bullet N \bullet N$ three-body system, 
we have binding because the attraction in the $K^- \bullet N$ sub-systems
is increased by a factor of $5/3$ when compared with the
$S=1$, $K^- \bullet N \bullet N$ three-body system. 
In particular we find
a binding energy $M - m_K-2m_N \in [-53.0,-14.2]$ MeV, 
and a decay width $\Gamma \in [13.6,28.3]$ MeV
to the $\pi \bullet \Sigma \bullet N$
and $\pi \bullet \Lambda \bullet N$ channels
\cite{Bicudo:2007fu}.
The complex pole of this resonance is comparable to the one we get for  
the $\Lambda(1405)$.

%
%
\begin{figure}[t]
\begin{picture}(230,100)(0,0)
\put(20,0){\put(-45,5){
\includegraphics[width=1.0\textwidth]{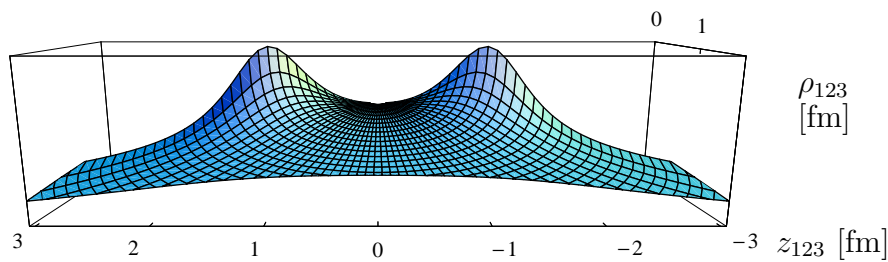}
}
\put(300,10){$z_{123}$ [fm]}
\put(308,70){$\rho_{123}$ }
\put(308,57){[fm]}
}
\end{picture}
\caption{3d perspective (colour online) of the antikaon wavefunction, assuming two 
adiabatically frozen nucleons at the distance of $r_{12}=2.5$ fm.
}
\label{double maximum}
\end{figure}

\section*{Acknowledgments}

PB thanks Avraham Gal, Eulogio Oset, Paola Gianotti, Marco Cardoso and George Rupp for suggestions and discussions.

\section{References}


\end{document}